\documentclass[twocolumn]{revtex4}
\usepackage{amsmath,amsfonts,amsthm,dsfont}

\def\Id{\textup{Id}}
\def\tr{\textup{tr}}

\def\gridspace{10}

\def\griddiagonal{5}
\def\griddot{\circle*{2}}

\begin{document}

\title{Decoupled Quantum Walks, models of the Klein-Gordon and wave equations}

\author{Pablo Arrighi}
\email{pablo.arrighi@imag.fr}
\affiliation{LIG, Universit\'e Joseph Fourier, Grenoble, France}
\affiliation{Universit\'e de Lyon, LIP, 46 all\'ee d'Italie, 69008 Lyon, France}  

\author{Stefano Facchini}
\email{stefano.facchini@imag.fr}
\affiliation{LIG, Universit\'e Joseph Fourier, Grenoble, France}

\date{\today}

\begin{abstract}
  Decoupling a vectorial PDE consists in solving the system for each
  component, thereby obtaining scalar PDEs that prescribe the
  evolution of each component independently.  We present a general
  approach to {\em decoupling} of Quantum Walks, again defined as a
  procedure to obtain an evolution law for each scalar component of
  the QW, in such a way that it does not depend on the other
  components.  In particular, the method is applied to show the
  relation between the Dirac (or Weyl) Quantum Walk in three space
  dimensions with (or without) mass term, and the Klein-Gordon (or
  wave) equation.
\end{abstract}

\maketitle

\section{Introduction} {\em Decoupling.} It is a common situation in Physics that a phenomenon is described by several magnitudes and a vectorial PDE governing them, but that we are interested in formulating a law for only one of those magnitudes, independently. This process is known as decoupling, when it succeeds it usually leads to a more complicated, higher-order scalar PDE for the magnitude under consideration (see for instance the well-known derivation of the Klein-Gordon equation from the Dirac equation \cite{Thaller,Bjorken}).\\
{\em Quantum Walks.} For the purpose of quantum simulation (on a quantum device) as envisioned by Feynman \cite{FeynmanQC}, or for the purpose of exploring the power and limits discrete models of physics, a great deal of efforts has gone into discretizing quantum physical phenomena. Most of these lead to a Quantum Walk (QW) model of the phenomena, i.e. a dynamics having the following features:
\begin{itemize}
\item The spacetime is a discrete grid;
\item The evolution is unitary; 
\item It is homogeneous, i.e. translation-invariant and time-independent;
\item It is causal, i.e. information propagates strictly at a bounded speed.
\end{itemize} 
In fact, QW models seem to prevail as the main discrete alternative to quantum physical PDEs. \\
{\em Decoupling Quantum Walks.} The main purpose of this paper is to formulate a notion of decoupling for QWs, and to provide a general technique for doing so. The technique is applied to the motivating example of the Dirac QW and its relation to the Klein-Gordon equation. \\
{\em QW model of the Klein-Gordon and wave equations.} QW models of Dirac equation have been extensively studied in the past two decades 
(\cite{BenziSucci,Bialynicki-Birula,MeyerQLGI,LoveBoghosian,
LapitskiDellarPalpacelliSucci,StrauchPhenomena,BoghosianTaylor1,DAriano,DAriano3D}). However the Dirac equation is first-order in time. Generally, it is not clear how PDEs with a second-order time
derivative can be modelled as QWs. However, given the particular relation between the Dirac equation and the Klein-Gordon equation, it should still be possible to obtain a QW model of the Klein-Gordon equation from the Dirac QW. This question was first tackled by \cite{IndiansDirac} in $(1+1)$-dimensions. A more precise treatment of the mass term, as well as space and time dependent generalizations, were given in \cite{MolfettaDebbasch}. Both papers proceed to a manual decoupling of the $(1+1)$-dimensional Dirac QW.\\
In this paper, we first proceed to a manual decoupling of the $(2+1)$-dimensional Dirac QW. We carry out explicit calculations in some details, and obtain an evolution law which admits the Klein-Gordon (KG) equation as the continuum limit. Next, we give a definition of the {\em decoupled form} of QW, and introduce a general procedure for
obtaining such a decoupling, via the minimal polynomial. Lastly, this procedure is applied to the $(3+1)$-dimensional QW. The decoupled form of this QW is shown to admits the square of the KG equation as the continuum limit.
As a final remark, a different discrete evolution law is shown, which has
the nice property of admitting the KG equation as the continuum limit. Unluckily, no underlying unitary QW could be found to have this evolution law as its decoupled form.

\section{Manual decoupling of the $(2+1)$ Dirac QW}  Let us consider the
following unitary operator
\begin{equation}
W_\varepsilon = e^{-i\varepsilon m\sigma^2}e^{-\varepsilon\sigma^1\partial_1}e^{-\varepsilon\sigma^3\partial_2}\label{eq:We}
\end{equation}
where the $\sigma^i$ are the Pauli matrices and $\varepsilon$ is a
real parameter representing the discretization step in time and in space.  That this is indeed a well-defined QW, whose continuum limit is 
\begin{align*}
i\partial_0 {\psi} &= (m\sigma^2 -i\sigma^1 \partial_1 -i\sigma^3 \partial_2)\,\psi
\end{align*}
the $2+1$-dimensional
Dirac equation for a particle with mass $m$, is proven in \cite{ArrighiDirac}.\\
On wavefunctions $\phi$ in $\mathbb{R}^2\rightarrow \mathbb{C}^2$ or $\varepsilon \mathbb{Z}^2\rightarrow \mathbb{C}^2$, the is QW naturally viewed as a discrete evolution $\phi \mapsto W_\varepsilon \phi$. On spacetime wavefunctions, the QW can be viewed as a constraint. In other words $\psi$ is a solution of the QW if and only if for all $t$,
\begin{align*}
\psi(t+\varepsilon) = W_\varepsilon \psi(t) 
\end{align*}
i.e. if it has been generated by the QW. In a more compact fashion, $\psi$ is a solution if and only if 
\begin{equation}
T_\varepsilon \psi = W_\varepsilon \psi \label{walk2d-eq}
\end{equation}
with  $T_\varepsilon = e^{\varepsilon\partial_0}$ the time translation operator, i.e. the operator verifying $(T_\varepsilon \psi)(t)=\psi(t+\varepsilon)$.
Explicitly, we can write the operator $W_\varepsilon$ as
\begin{equation*}
  W_\varepsilon =
  \begin{pmatrix}
    W_{ll} & W_{lr} \\
    W_{rl} & W_{rr}
  \end{pmatrix}
\end{equation*}
where, according to Eq. \eqref{eq:We}:
\begin{equation*}
\begin{split}
  W_{ll} &= \frac{c-s}2 \tau_x\tau_y + \frac{c+s}2 \tau_x^{-1}\tau_y, \\
  W_{lr} &= \frac{c-s}2 \tau_x\tau_y^{-1} - \frac{c+s}2 \tau_x^{-1}\tau_y^{-1}, \\
  W_{rl} &= \frac{c+s}2 \tau_x\tau_y - \frac{c-s}2 \tau_x^{-1}\tau_y, \\
  W_{rr} &= \frac{c+s}2 \tau_x\tau_y^{-1} + \frac{c-s}2 \tau_x^{-1}\tau_y^{-1}
\end{split}
\end{equation*}
with
\begin{equation*}
  c = \cos(\varepsilon m), \quad s = \sin(\varepsilon m), \quad \tau_x = e^{-\varepsilon \partial_1}, \quad \tau_y = e^{-\varepsilon \partial_2}.
\end{equation*}
Then, equation \eqref{walk2d-eq} can be written as two coupled equations
for $\psi_l$ and $\psi_r$:
\begin{align*}
  T_\varepsilon\psi_l = \left( W_{ll}\psi_l + W_{lr}\psi_r \right), \\
  T_\varepsilon\psi_r = \left( W_{rl}\psi_l + W_{rr}\psi_r \right).
\end{align*}
Multiplying the first by $W_{rl}$ we have
\begin{equation*}
  T_\varepsilon W_{rl}\psi_l = \left( W_{ll}W_{rr}\psi_l + W_{lr}W_{rl}\psi_r \right)
\end{equation*}
and replacing $W_{rl}\psi_l$ by $T_\varepsilon\psi_r - W_{rr}\psi_r$ (i.e. the
second equation) we have
\begin{equation*}
  \left[T_\varepsilon^2-(W_{ll}+W_{rr})T_\varepsilon + (W_{ll}W_{rr}-W_{lr}W_{rl})\right]\psi_r = 0.
\end{equation*}
We recognize here the trace and the determinant of $W_\varepsilon$.
Notice that 
\begin{align*}
\det(W_\varepsilon)&=\det(e^{-i\varepsilon m\sigma^2})\det(e^{-\varepsilon\sigma^1\partial_1})\det(e^{-\varepsilon\sigma^3\partial_2})\\
&=e^{\tr(-i\varepsilon m\sigma^2)}e^{\tr(-\varepsilon\sigma^1\partial_1)}e^{\tr(-\varepsilon\sigma^3\partial_2)})=1
\end{align*}
Therefore we reach:
\begin{equation}
  \left[T_\varepsilon^2-\tr(W_\varepsilon)T_\varepsilon + 1\right]\psi_r = 0 \label{kg-walk-2d}
\end{equation}
(the same equation holds for the $\psi_l$ component.)  This is the decoupled form of the Dirac QW \cite{ArrighiDirac}. It can be interpreted as an evolution law which prescribes $\psi_r$ at the next time step as a function of $\psi_r$ at the present and past time steps, i.e.
$$\psi_r(t+\varepsilon)=\tr(W_\varepsilon)\psi_r(t)- \psi_r(t-\varepsilon).$$
Explicitly:
\begin{equation}
  \tr(W_\varepsilon) = \frac{c-s}2(\tau_x\tau_y+\tau_x^{-1}\tau_y^{-1})+\frac{c+s}2(\tau_x^{-1}\tau_y+\tau_x\tau_y^{-1}) \label{kg-trace-2d}
\end{equation}
Let us call $P_\varepsilon$ the LHS of Eq. \eqref{kg-walk-2d}. 
We show now that the continuum limit of the decoupled form is the
KG equation, i.e. that:
\begin{equation*}
  \lim_{\varepsilon \to 0} \frac{p_\varepsilon}{\varepsilon^2} = \partial_0^2 - \partial_1^2 - \partial_2^2 + m^2
\end{equation*}
We use the following Taylor expansion in $\varepsilon$ to second order:
\begin{align*}
  T_\varepsilon &= 1 + \varepsilon\partial_0 + \frac{\varepsilon^2}2\partial_0^2 + O(\varepsilon^3)\\
  T^2_\varepsilon &= 1 + 2\varepsilon\partial_0 + 2\varepsilon^2\partial_0^2 + O(\varepsilon^3)\\
  e^{-i\varepsilon m\sigma^2} &= \Id-i\varepsilon m \sigma^2 - \frac{m^2\varepsilon^2}2 + O(\varepsilon^3)\\
  e^{-\varepsilon\sigma^1\partial_1} &= \Id-\varepsilon \sigma^1\partial_1 + \frac{\varepsilon^2}2\partial_1^2 + O(\varepsilon^3)\\
  e^{-\varepsilon\sigma^3\partial_2} &= \Id-\varepsilon \sigma^3\partial_2 + \frac{\varepsilon^2}2\partial_2^2 + O(\varepsilon^3).
\end{align*}
Inserting these in $\tau_\varepsilon$ and collecting terms with similar order indeed gives
\begin{equation*}
  p_\varepsilon = \varepsilon^2(\partial_0^2 - \partial_1^2 - \partial_2^2 + m^2) + O(\varepsilon^3)
\end{equation*}
and the statement follows immediately.\\
An important remark is that whilst the decoupled form Eq. \eqref{kg-walk-2d} can be interpreted as an evolution law, this evolution law needs two time slices $\psi(0)$ and $\psi(-\varepsilon)$ as initial conditions, which could be unrelated. Thus, the spacetime wavefunctions that are solutions of the decoupled form of the QW are not necessarily solutions of the QW. Some information is lost about the dynamics through decoupling. In fact, one can easily construct a $W_\varepsilon'$ which differs form $W_\varepsilon$ but has the same trace and determinant and thus the same decoupling --- for instance by bringing the mass term of Eq. \eqref{eq:We} to the front.

\section{Decoupling Quantum Walks}  Given a QW defined by a unitary operator $W_\varepsilon$ and having solutions spacetime wavefunctions $\psi$ in 
$\mathbb{R}^{n+1}\rightarrow \mathbb{C}^{d}$ (or $\varepsilon\mathbb{Z}^{n+1}\rightarrow \mathbb{C}^{d}$), we can apply
the same procedure as in the previous section, i.e. proceed by
``Gaussian elimination'' in order to find independent equations for
each component of $\psi$.\\
More formally, a {\em decoupled form} of the QW $W_\varepsilon$ is an equation
\begin{equation}
  \psi(t+\varepsilon s) = \sum_{k<s} a_k \psi(t+\varepsilon k)\label{eq:decoupled}
\end{equation}
where the $a_k$ are operators over scalar wavefunctions $\psi_i$ in 
$\mathbb{R}^{n}\rightarrow \mathbb{C}$ (here naturally extended to $\mathbb{R}^{n}\rightarrow \mathbb{C}^d$), and such that the equation holds on every $\psi$ a solution of $W_\varepsilon$.\\
In general, we want minimize $s$. We will now provide a general procedure for doing so. 
Let $\psi$ be a solution of $W_\varepsilon$, Eq. \eqref{eq:decoupled} becomes:
\begin{equation*}
  \left(W_\varepsilon^s-\sum_{k<s}a_kW_\varepsilon^k\right)\psi(t) = 0
\end{equation*}
\begin{equation*}
\textrm{or}\qquad  p(W_\varepsilon)\psi(t) = 0\qquad \textrm{with} \qquad p(\lambda) = \lambda^s - \sum_{k<s}a_k\lambda^k.
\end{equation*}
From this expression we see that our goal is to find a monic
polynomial $p(\lambda)$ of minimal degree (i.e. minimal $s$) such that
$p(W_\varepsilon)=0$, in other words, the {\em minimal polynomial} of
$W_\varepsilon$ \cite{Lang}. Once this polynomial is found, we can also write
\begin{equation}
  p(T_\varepsilon)\psi = 0 \label{eq-gen-dec}
\end{equation}
because $T_\varepsilon\psi=W_\varepsilon\psi$ holds by definition of
the solutions of the $W_\varepsilon$. Equation \eqref{eq-gen-dec} is the {\em minimal decoupled form} of the QW.\\ 
In order to reach an explicit form, recall that it is well-known from linear algebra (see \cite{Lang}) that the general form of the minimal polynomial is
\begin{equation*}
  p(\lambda) = \prod_j (\lambda-\lambda_j),
\end{equation*}
with $\lambda_j$ ranging over the {\em distinct} eigenvalues of
$W_\varepsilon$.  An important remark is that if all eigenvalues of $W_\varepsilon$ are distinct, then the minimal polynomial coincides with the {\em characteristic
  polynomial}:
\begin{equation*}
  \chi_{W_\varepsilon}(\lambda) = \det(\lambda\Id-W_\varepsilon).
\end{equation*}
This form is particularly convenient, as it allows to obtain the
decoupled form of $W_\varepsilon$ from a determinant, instead of an eigenvalue computation
(which is potentially a hard task).

Notice that we may then obtain the continuum limit of the decoupled form as the limit
\begin{equation*}
  \lim_{\varepsilon\to 0}\frac{p(T_\varepsilon)}{\varepsilon^s},
\end{equation*}
if it exists. 

\section{Decoupling of the $(3+1)$ Dirac QW}
We now apply the above procedure to the following QW:
\begin{equation*}
  W_\varepsilon = e^{-i\varepsilon m\beta}e^{-\varepsilon \alpha^1\partial_1}e^{-\varepsilon \alpha^2\partial_2}e^{-\varepsilon \alpha^3\partial_3}
\end{equation*}
%or, in Fourier space,
%\begin{equation}
%  W_\varepsilon = e^{-i\varepsilon m\beta}e^{-i\varepsilon \alpha^1 k_x}e^{-i\varepsilon \alpha^2 k_y}e^{-i\varepsilon \alpha^3 k_z}.
%\end{equation}
where $\alpha^i=\sigma^3\otimes\sigma^i, \beta=\sigma^2\otimes\sigma^0$ are a four-dimensional representation of the
Dirac matrices. That this is indeed a well-defined QW, whose continuum limit is 
\begin{align*}
i\partial_0 {\psi} &= (m\beta +i \alpha^1\partial_1+i \alpha^2\partial_2+i \alpha^3\partial_3)\,{\psi}
\end{align*}
the $3+1$-dimensional
Dirac equation for a particle with mass $m$, is proven in \cite{ArrighiDirac}.\\
Since $W_\varepsilon$ has four distinct eigenvalue (as can be proved
numerically), its minimal decoupling can be obtained through its characteristic polynomial.
%for a $4\times 4$ matrix $A$ is
%\begin{equation}
%\tau_A(\lambda) = \lambda^4 - \tr(A)\lambda^3 + \frac12[\tr(A)^2 - \tr(A^2)]\lambda^2 - \frac16[\tr(A)^3 - 3\tr(A^2)\tr(A) + 2\tr(A^3)]\lambda + \det(A).
%\end{equation}
Our decoupled form then reads
\begin{equation*}
  \chi_{W_\varepsilon}(T_\varepsilon)\psi = 0.
\end{equation*}
with the characteristic polynomial \cite{Lang}:
\begin{equation*}
  \begin{split}
    \chi_{W_\varepsilon}(T_\varepsilon) = T_\varepsilon^4 - \tr(W_\varepsilon)T_\varepsilon^3 + \frac12[\tr(W_\varepsilon)^2 - \tr(W_\varepsilon^2)]T_\varepsilon^2 - \\
    \frac16[\tr(W_\varepsilon)^3 - 3\tr(W_\varepsilon^2)\tr(W_\varepsilon) + 2\tr(W_\varepsilon^3)]T_\varepsilon + 1
  \end{split}
\end{equation*}
We can prove that the continuum limit of the decoupled 3D QW is the
(square of) KG equation
\begin{equation}
  \lim_{\varepsilon \to 0} \frac{\chi_{W_\varepsilon}(T_\varepsilon)}{\varepsilon^4} = (\Box + m^2)^2 \label{eq-theorem}
\end{equation}
where $\Box=\partial_0^2-\partial_1^2-\partial_2^2-\partial_3^2$.
In order to prove this statement, we shall first prove that for any matrices $\alpha^i \in \mathcal{M}_d(\mathbb{C})$, and scalars $A_i$, the following
holds
\begin{equation}\label{eq:lemma}
  \det\left(\Id - \prod_i e^{-\varepsilon \alpha^i A_i}\right) = \varepsilon^d \det(\alpha \cdot A) + O(\varepsilon^{d+1})
\end{equation}
In fact we have
\begin{align*}
  & \det\left(\Id - \prod_i e^{-\varepsilon \alpha^i A_i}\right) \\
  = & \det\left(\Id - e^{-\varepsilon \sum_i \alpha^i A_i + O(\varepsilon^2)}\right) = \det\left(\Id - e^{-\varepsilon \alpha \cdot A + O(\varepsilon^2)}\right) = \\
  = & \det\left(\Id - (\Id - \varepsilon \alpha \cdot A + O(\varepsilon^2))\right) = \det (\varepsilon \alpha \cdot A + O(\varepsilon^2)) = \\
  = & \varepsilon^d \det(\alpha \cdot A + O(\varepsilon)) = \varepsilon^d \det(\alpha \cdot A) + O(\varepsilon^{d+1})
\end{align*}
where the first equality is obtained by the BCH formula.

We now proceed to calculate the limit of equation \eqref{eq-theorem}.  Using $T_\varepsilon=e^{\varepsilon\partial_0}$ and the definition of $W_\varepsilon$, the
expression undergoing the limit can be rewritten as
\begin{align*}
  \frac{\chi_{W_\varepsilon}(T_\varepsilon)}{\varepsilon^4} = \frac{1}{\varepsilon^4}\det\left(e^{\varepsilon\partial_0} \Id - e^{-\varepsilon i m \beta} \prod_i e^{-\varepsilon \alpha^i\partial_i} \right) \\
  = \frac{1}{\varepsilon^4}\det(e^{\varepsilon\partial_0}\Id) \det\left(\Id - e^{-\varepsilon\partial_0}e^{-\varepsilon i m \beta} \prod_i e^{-\varepsilon \alpha^i\partial_i} \right) \\
  = \det(e^{\varepsilon\partial_0}\Id) \left[\det(\partial_0 + i m \beta + \alpha \cdot \nabla) + O(\varepsilon)\right]
\end{align*}
where in the third equality we used equation \eqref{eq:lemma}. Taking now the
limit $\varepsilon \to 0$ this leads to
\begin{align*}
  \lim_{\varepsilon \to 0} \frac{\chi_{W_\varepsilon}(T_\varepsilon)}{\varepsilon^4} =& \det(\partial_0 + i m \beta + \alpha \cdot \nabla) \\
  =& \det\begin{pmatrix}
    \partial_0 \Id +\sigma \cdot \nabla  && m\Id \\
     -m\Id  && \partial_0\Id-\sigma \cdot \nabla
  \end{pmatrix} \\
  =& \det((\partial_0^2 + m^2 - \nabla^2)\Id) 
\end{align*}
as obtained by evaluating the determinant partially \cite{Silvester}. We then recognize
$(\Box + m^2)^2$ which completes the proof.\\
We notice that, while the usual decoupling
of the Dirac equation gives the Klein-Gordon equation, the decoupling
of the Dirac QW can only give the square of the Klein-Gordon
operator. This is somehow unexpected, but it is entailed by the fact
that the Dirac QW evolution operator $W_\varepsilon$ breaks the
eigenvalue degeneracy of the continuous Dirac evolution, leading to a
minimal polynomial of degree four. However, the decoupling procedure does not introduce spurious plane wave solutions. Indeed, the differential operators $\Box + m^2$ and $(\Box + m^2)^2$ yield differential equations having the same solutions when $\psi(t, \mathbf{x})=e^{i\omega t - i\mathbf{k}.\mathbf{x}}$, i.e.
\begin{align*}
  & (\Box + m^2)^2 \psi(t, \mathbf{x}) = 0 \Longleftrightarrow (\Box + m^2) \psi(t, \mathbf{x}) = 0
\end{align*}
since $(-\omega^2+k^2+m^2)^2 = 0$ if and only if $-\omega^2+k^2+m^2 = 0$. In other words, the dispersion relation is unchanged.

\section{Decoupling of the $(3+1)$ Weyl QW}
We now apply the decoupling procedure to the following QW:
\begin{equation*}
  W_\varepsilon = e^{-\varepsilon \sigma^1\partial_1}e^{-\varepsilon \sigma^2\partial_2}e^{-\varepsilon \sigma^3\partial_3}
\end{equation*}
That this is indeed a well-defined QW, whose continuum limit is the Weyl equation:
\begin{equation*}
\partial_0 \psi = -\sigma^1 \partial_1 -\sigma^2 \partial_2 -\sigma^3 \partial_3
\end{equation*}
directly follows from \cite{Bialynicki-Birula,ArrighiDirac}.\\
Since in this case the coin space has dimension two, the decoupled
form is again obtained through the characteristic polynomial, i.e.
\begin{equation*}
\chi_{W_\varepsilon}(T_\varepsilon) = T_\varepsilon^2 - \tr(W_\varepsilon)T_\varepsilon + 1
\end{equation*}
and making use of equation \eqref{eq:lemma} as in the previous section we reach
\begin{align*}
  \lim_{\varepsilon \to 0} \frac{\chi_{W_\varepsilon}(T_\varepsilon)}{\varepsilon^2} &= \det(\partial_0 + \sigma^1\partial_1 + \sigma^2\partial_2 + \sigma^3\partial_3) \\
&= \partial_0^2 - \partial_1^2 - \partial_2^2 - \partial_3^2
\end{align*}
i.e. the wave equation. Thus, if there is no mass, decoupling can yields directly the wave equation instead of its square --- even in $(3+1)$-dimensions.

\section{A conjecture} An interesting question is to look for
some unitary QW whose decoupling gives the KG in the limit, without
the spurious squaring. A direct extrapolation from equations
\eqref{kg-walk-2d} and \eqref{kg-trace-2d} leads for instance to a
discrete evolution of the form
 \begin{equation*}
  \left\{T_\varepsilon^2-T_\varepsilon(A_++A_-)-I\right\}\psi = 0 \label{extrapolation-eq}
\end{equation*}
where
\begin{equation*}
  \begin{split}
    A_+ = \frac14 e^{-im\varepsilon}(\tau_x\tau_y\tau_z + \tau_x\tau_y^{-1}\tau_z^{-1} +\\
    + \tau_x^{-1}\tau_y\tau_z^{-1} + \tau_x^{-1}\tau_y^{-1}\tau_z)
  \end{split}
\end{equation*}
and
\begin{equation*}
  \begin{split}
    A_- = \frac14 e^{im\varepsilon}(\tau_x\tau_y\tau_z^{-1} + \tau_x\tau_y^{-1}\tau_z +\\
    + \tau_x^{-1}\tau_y\tau_z + \tau_x^{-1}\tau_y^{-1}\tau_z^{-1})
  \end{split}.
\end{equation*}

A graphical representation of $A_\pm$ is as follows:

\begin{equation*}
A_+ = \frac{e^{-i\varepsilon m}}{4}\left(
\setlength{\unitlength}{1mm}
\begin{picture}(20, 15)
  \linethickness{0.5pt}

  % dots
  \put(2,-5){\griddot}
  \put(12,5){\griddot}
  \put(7,10){\griddot}
  \put(17,0){\griddot}

  % straight lines
  \multiput(2, -5)(5, 5){2}{
    \multiput(0, 0)(0, \gridspace){2}{\line(1, 0){\gridspace}}
    \multiput(0, 0)(\gridspace, 0){2}{\line(0, 1){\gridspace}}
  }

  % diagonal lines
  \multiput(2, -5)(0, \gridspace){2}{
    \multiput(0, 0)(\gridspace, 0){2}{\line(1, 1){\griddiagonal}}
  }
\end{picture}
\right)
\end{equation*}
\begin{equation*}
A_- = \frac{e^{+i\varepsilon m}}{4}\left(
\setlength{\unitlength}{1mm}
\begin{picture}(20, 15)
  \linethickness{0.5pt}

  % dots
  \put(12,-5){\griddot}
  \put(2,5){\griddot}
  \put(17,10){\griddot}
  \put(7,0){\griddot}

  % straight lines
  \multiput(2, -5)(5, 5){2}{
    \multiput(0, 0)(0, \gridspace){2}{\line(1, 0){\gridspace}}
    \multiput(0, 0)(\gridspace, 0){2}{\line(0, 1){\gridspace}}
  }

  % diagonal lines
  \multiput(2, -5)(0, \gridspace){2}{
    \multiput(0, 0)(\gridspace, 0){2}{\line(1, 1){\griddiagonal}}
  }
\end{picture}
\right).
\end{equation*}

The continuum limit of this expression is indeed the KG equation in
three dimensions
\begin{equation*}
  \lim_{\varepsilon\to 0}\frac{T_\varepsilon^2-T_\varepsilon(A_++A_-)-I}{\varepsilon^2} = \Box + m^2
\end{equation*}
without squaring. This can be proved by a Taylor expansions of the involved terms, along the lines of the derivation of the continuum limit for the $(2+1)$-dimensional decoupled QW. But unluckily we could not find any unitary QW whose
decoupling is equation \eqref{extrapolation-eq}. We conjecture that it does not exist. Although not being a
proof, one could argue that such QW would probably have the Dirac
equation as continuum limit, providing then a two-dimensional
representation of the Dirac matrices, which is known to be impossible.

\section{Conclusions} In this paper, we provided a general approach to decoupling Quantum Walks. In particular, we provided a procedure to obtain the minimal decoupled form of a Quantum Walk from its eigenvalues, via the minimal polynomial. When the Quantum Walk has distinct eigenvalues, the procedure simplifies to a convenient formula expressed in terms of its characteristic polynomial.\\
We motivated our analysis via the manual derivation of the minimal decoupled
form of the $(2+1)$-dimensional Dirac QW with mass. As expected, the continuum limit of this minimal decoupled form is the KG equation.\\
We then applied our formula to the derivation of the minimal decoupled
form of the $(3+1)$-dimensional Dirac QW with mass. Surprisingly, the continuum limit of this minimal decoupled form turns out to be the square of the KG equation. This is a consequence of the eigenvalue degeneracy-breaking of the Dirac QW with respect to the original Dirac equation.\\
Finally we applied our formula to the derivation of the minimal decoupled
form of the $(3+1)$-dimensional Weyl QW. Reassuringly, the continuum limit of this minimal decoupled form turned out to be the wave equation.\\
In \cite{Bialynicki-Birula,ArrighiDirac} the authors give methods to make the first-order, free fields equations of the fundamental spin $1/2$ and spin $1$ particles (and symmetric hyperbolic systems in general) into Quantum Walks. This paper gives a general method which can be applied to recover the Klein-Gordon and the Wave equations from those. It thus seems that Quantum Walks provide a discrete formulation of the free fields evolution laws of the fundamental particles of the standard model, which is a promising route for investigation.

\section*{Acknowledgments}
We would like to thank Vincent Nesme and Marcelo Forets for several
helpful discussions.  This work has been funded by the
ANR-10-JCJC-0208 CausaQ grant.

\bibliography{../../Bibliography/biblio}

\end{document}